\pdfoutput=1 

\documentclass[preprint,superscriptaddress,nofootinbib,longbibliography]{revtex4-1}
\usepackage[T1]{fontenc}
\usepackage{amsmath}
\usepackage{amssymb}
\usepackage{hyperref} 
\usepackage{footnote}

\providecommand{\tabularnewline}{\\}

\begin{document}
\title{Baryogenesis in the Standard Model and its Supersymmetric Extension}

\newcommand{\affUFABC}{Centro de Ci\^encias Naturais e Humanas\;\;\\
		Universidade Federal do ABC, 09.210-170,
		Santo Andr\'e, SP, Brazil}

\author{Chee Sheng Fong}
\email{sheng.fong@ufabc.edu.br}
\affiliation{\affUFABC}

\begin{abstract}
In this work, we classify all the effective $U(1)$ symmetries and their associated Noether charges in the Standard Model (SM) and its minimal supersymmetric extension (MSSM) from the highest scale after inflation down to the weak scale. We then demonstrate that the discovery of the violation of baryon minus lepton number ($B-L$) which pinpoints to its violation in primordial Universe at any cosmic temperature above $30$ TeV will open up a new window of baryogenesis in these effective charges above this scale. 
While the fast SM baryon number violation in the early Universe could be the first piece to solve the puzzle of cosmic baryon asymmetry, $(B-L)$ violation could be the second important piece. In the background of expanding Universe, there is ample opportunity for out-of-equilibrium processes to generate an asymmetry in the numerous effective charges in the SM or the MSSM, making a baryon asymmetric Universe almost \emph{unavoidable}. 
Finally we provide examples in the SM and the MSSM where baryogenesis can proceed through out-of-equilibrium dynamics without explicitly breaking baryon nor lepton number.

\end{abstract}

\maketitle
\flushbottom

\section{\label{sec:intro}Introduction}

Symmetry has been a guiding principle in physics. In the Standard
Model (SM), the gauge symmetry $SU(3)_{c}\times SU(2)_{L}\times U(1)_{Y}$
fixes all the possible interactions among the fields. However, once
one consider the same theory in an expanding Universe, new effective symmetries can arise if the rates of certain interactions
are slower than the Hubble rate, i.e. out of thermal equilibrium.
For instance, while baryon-number-violating interactions due to $SU(2)_{L}$ instanton-induced operators are suppressed to an unobservable rate today \cite{tHooft:1976rip}, they were in thermal equilibrium for
cosmic temperature above the weak scale \cite{Kuzmin:1985mm}. This
baryon number violation in the primordial universe, among others, could
be a key ingredient to understand the observed cosmic baryon asymmetry
represented by the baryon number density asymmetry over cosmic entropic density
$Y_{\Delta B}\sim9\times10^{-11}$ \cite{Zyla:2020zbs}. 

Symmetry in the context of baryogenesis is a double-edged sword. On the one hand, if a symmetry remains exact, no asymmetry can be generated in the associated Noether charge. On the other hand, it can protect an asymmetry generated from being erased. This was first elucidated by Weinberg \cite{Weinberg:1980bf}
that as long as there is a linear combination of baryon and lepton
number $B+aL$ that is conserved, a baryon asymmetry is preserved
even if baryon-number-violating interactions are in equilibrium. 
The use of other types of symmetry as protective mechanism was further explored in the SM \cite{Campbell:1992jd,Cline:1993vv,Cline:1993bd},
in its extension \cite{Antaramian:1993nt,Sierra:2013kba}, in its
supersymmetric extension \cite{Ibanez:1992aj} and was generalized in \cite{Fong:2015vna}.

In this work, we will categorize all the effective symmetries and their associated Noether charges in the SM and the Minimal Supersymmetric Standard Model (MSSM), from the highest temperature after reheating $T\sim10^{16}$ GeV down to the weak scale. 
Making use of $(B-L)$-violating interactions which are \emph{in equilibrium}, we show that asymmetries can be generated in any of the effective charges through \emph{out-of-equilibrium} processes which can, but \emph{do not} have to violate baryon nor lepton number.

\section{\label{sec:generality}Generality}

Here we give a brief review of the formalism discussed in \cite{Fong:2015vna}
which will be used in this work and also to fix the notations. For
any particle species $i$ that distinguishes from its antiparticle
$\overline{i}$ i.e. a ``complex particle'',
its number density asymmetry can be defined as
$n_{\Delta i}\equiv n_{i}-n_{\overline{i}}$ where $n_{i}$ is the
number density of particle $i$. This is equivalent to its charge
density if we assign the particle (antiparticle) a charge $1(-1)$
under a global $U(1)_{i}$. Considering $r$ species of complex particles $i$, we can define $r$ number of such global $U(1)_{i}$ preserved by their kinetic terms (which include possible gauge interactions).

Next, let us put the system in an expanding Universe. Assuming fast gauge interactions are able to thermalize the particles, a common temperature $T$ can be defined at each moment. To scale out the effect due to cosmic expansion, we define the number asymmetry of the particle $i$ as
\begin{eqnarray}
Y_{\Delta i} & = & \frac{n_{\Delta i}}{s},
\end{eqnarray}
where $s=\frac{2\pi}{45}g_{\star}T{{}^3}$ is the cosmic entropic density with $g_{\star}$ the total relativistic degrees of freedom.
From $r$ number of $U(1)_{i}$,
we can form any other $r$ linear combinations $U(1)_{x}$. 
A convenient basis is such that as the expanding
Universe is cooling down, $U(1)_{x}$ is subsequently broken by interactions which get into thermal equilibrium. 

The source of $U(1)_{x}$ can be due to additional interactions like the Yukawa interactions or those from instanton-induced
effective operators due to the Bell-Jackiw anomaly \cite{tHooft:1976rip}.
To determine if a $U(1)_{x}$ is preserved by the latter type of interactions,
we can look at the anomaly coefficient of the triangle
diagram of the type $U(1)_{x}-SU(N)-SU(N)$ defined as
\begin{eqnarray}
A_{xNN} & \equiv & \sum_{i}c\left(R_{i}\right)g_{i}q_{i}^{x},\label{eq:anomaly_coeff}
\end{eqnarray}
where the sum is over particle $i$ with degeneracy $g_{i}$, charge
$q_{i}^{x}$ under $U(1)_{x}$, and representation $R_{i}$ under
$SU(N\geq 2)$ gauge group with $c\left(R_{i}\right)=\frac{1}{2}$
in the fundamental representation and $c_{2}\left(R_{i}\right)=N$
in the adjoint representation. In general, each fermion $i$ with
representation $R_{i}$ under $SU(N)$ will contribute proportionally
to $c\left(R_{i}\right)$ to the $SU(N)$ instanton-induced effective
operator~\footnote{Due to the convention $c\left(R_{i}\right)=\frac{1}{2}$ for fundamental
representation, the factor of 2 is included such that the field enters
in integer number.}
\begin{eqnarray}
{\cal O}_{SU(N)} & \sim & \prod_{i}\Psi_{i}^{2g_{i}c\left(R_{i}\right)},\label{eq:instanton_induced_operators}
\end{eqnarray}
where the product is over all left-handed chiral fields $\Psi_{i}$ with
nontrivial representation $R_{i}$ under $SU(N)$. Notice that if $A_{xNN}=0$, $U(1)_{x}$ is conserved by the operator.

If a system possesses a $U(1)_{x}$, its charge can be written as
\begin{eqnarray}
Y_{\Delta x} & = & \sum_{i}q_{i}^{x}Y_{\Delta i},
\end{eqnarray}
where $i$ sum over all particle species with charge $q_{i}^{x}$ under $U(1)_{x}$. Assuming equilibrium phase space distribution for particle $i$ characterized by a common temperature $T$, its energy $E_{i}$ and chemical potential $\mu_{i}\ll T$, the formula above can be inverted as \cite{Fong:2015vna} 
\begin{eqnarray}
Y_{\Delta i} & = & \sum_{x}g_{i}\zeta_{i}\sum_{y}q_{i}^{y}\left(J^{-1}\right)_{yx}Y_{\Delta x},\label{eq:general_number_asymmetry}
\end{eqnarray}
where $\zeta_{i}=1(2)$ for $i$ a massless fermion (boson)~\footnote{For a particle with mass $m_i$, $\zeta_{i}=\frac{6}{\pi^{2}}\int_{m_{i}/T}^{\infty}dxx\sqrt{x^{2}-m_{i}^{2}/T^{2}}\frac{e^{x}}{\left(e^{x}\pm\xi_{i}\right)^2}$ where $\xi_{i}=1\left(-1\right)$ for $i$ a fermion (boson).} 
and $J$ is a symmetric matrix in charge space defined as 
\begin{eqnarray}
J_{xy} & \equiv & \sum_{i}g_{i}\zeta_{i}q_{i}^{x}q_{i}^{y}.
\label{eq:J matrix}
\end{eqnarray}
From eq. (\ref{eq:general_number_asymmetry}), we can construct the baryonic charge as 
\begin{eqnarray}
Y_{\Delta B} & = & \sum_{i}q_{i}^{B}Y_{\Delta i}=\sum_{x}\sum_{y}J_{By}\left(J^{-1}\right)_{yx}Y_{\Delta x}.\label{eq:baryon_asymmetry_general}
\end{eqnarray}
It is apparent that the cosmic baryon asymmetry is
proportional to the $U(1)_{x}$ charges of the system. Next, our goal
is to characterize all the $U(1)_{x}$ in the SM and MSSM.

\section{\label{sec:SM}The Standard Model}

Before the electroweak (EW) symmetry breaking, the SM kinetic terms
respect a total of 16 $U(1)_{\Psi_{j}}$, each corresponds to the individual
field rotation of the 15 fermionic fields and a scalar Higgs
$SU(2)_{L}$ doublet $H$: $\Psi_{j}\to e^{iq_{\Psi_j}\phi}\Psi_{j}$ where
$\Psi=\left\{ Q_{\alpha},U_{\alpha},D_{\alpha},\ell_{\alpha},E_{\alpha},H\right\} $
with family index $\alpha=1,2,3$. Here $Q_{\alpha},\ell_{\alpha}$
are respectively the quark and lepton $SU(2)_{L}$ doublets while $U_{\alpha},D_{\alpha},E_{\alpha}$
are respectively the up-type quark, down-type quark and lepton $SU(2)_{L}$ singlets.~\footnote{We will also denote $\left\{ U_{1},U_{2},U_{3}\right\} =\left\{ u,c,t\right\} $,
$\left\{ D_{1},D_{2},D_{3}\right\} =\left\{ d,s,b\right\} $ and $\left\{ E_{1},E_{2},E_{3}\right\} =\left\{ e,\mu,\tau\right\} $.} 
Since all the parameters of the SM have been measured, we can choose
the following convenient basis according to the order in which the $U(1)_{x}$ symmetries are subsequently broken at $T_x$ as we go down in the cosmic temperature with
\begin{eqnarray}
x & = & \left\{ t,u,B,u-b,\text{\ensuremath{\tau}},u-c,\mu,B_{3}-B_{2},u-s,B_{3}+B_{2}-2B_{1},u-d,e,B/3-L_{\alpha},Y\right\} .\label{eq:SM_global_symmetries}
\end{eqnarray}
$U(1)_{B_{\alpha}}$ correspond to baryon flavor symmetries with $q^{B_\alpha}_{Q_\alpha,U_\alpha,D_\alpha}=1/3$ and $U(1)_{B}\equiv U(1)_{B_{1}+B_{2}+B_{3}}$ is the total baryon number.
$U(1)_{B/3-L_{\alpha}}$ are the linear combination of $B/3$ and lepton flavor symmetries $U(1)_{L_{\alpha}}$ with $q^{L_\alpha}_{\ell_\alpha,E_\alpha}=1$.
The rest of the global charges are normalized to 1. 

$U(1)_{Y}$ is the hypercharge gauge symmetry which is only broken at $T_{\textrm{EW}}\sim160\,\textrm{GeV}$ \cite{DOnofrio:2014rug} while the rest are effective (global) symmetries which are broken at $T_t \sim 10^{15}$ GeV, $T_u \sim 2 \times 10^{13}$ GeV, $T_B \sim 2 \times 10^{12}$ GeV, $T_\tau \sim 4 \times 10^{11}$ GeV, $T_{u-b} \sim 3 \times 10^{11}$ GeV, $T_{u-c} \sim 2 \times 10^{10}$ GeV, $T_\mu \sim 10^9$ GeV, $T_{B_3-B_2} \sim 9 \times 10^8$ GeV, $T_{u-s} \sim 3 \times 10^8$ GeV, $T_{B_3+B_2-2B_1} \sim 10^7$ GeV, $T_{u-d} \sim 2 \times 10^6$ GeV and $T_e \sim 3 \times 10^4$ GeV. The estimations of $T_{x}$ are explained in Appendix \ref{app:symmetries_T}. Once broken at $T_x$, the symmetries are not restored again with the exception of $U(1)_B$. From eq. \eqref{eq:instanton_induced_operators}, we have the $SU(2)_{L}$ instanton-induced operator ${\cal O}_{SU(2)_{L}}\sim\prod_{\alpha}\left(Q\ell\ell\ell\right)_{\alpha}$ 
which violates $U(1)_{B}$ and the associated processes are in thermal equilibrium
\cite{Kuzmin:1985mm} from $T_{B}\sim2\times10^{12}$ GeV \cite{Garbrecht:2014kda}
down to $T_{B-}\sim130$ GeV \cite{DOnofrio:2014rug}. This is the source of baryon number violation for EW baryogenesis \cite{Morrissey:2012db}, though in the SM, two other Sakharov's conditions for baryogenesis \cite{Sakharov:1967dj}, sufficient C and CP violation \cite{Gavela:1993ts,Huet:1994jb}, and sufficiently out-of-equilibrium processes \cite{Kajantie:1996qd} are not fulfilled. 

Among the global symmetries, only $U(1)_{B/3-L_{\alpha}}$ (or any linear combinations) are exact: they are conserved by all the Yukawa interactions and free from $SU(2)_{L}$ and $SU(3)_{c}$ mixed anomalies. At temperature below $T_{e}\sim 30\,\textrm{TeV}$, only $U(1)_{Y}$ and $U(1)_{B/3-L_{\alpha}}$ remain conserved and from eq. (\ref{eq:baryon_asymmetry_general}),
we have
\begin{eqnarray}
Y_{\Delta B} & = & c_{B\left(B-L\right)}Y_{\Delta(B-L)}+c_{BY}Y_{\Delta Y},
\label{eq:baryon_asymmetry}
\end{eqnarray}
where $Y_{\Delta(B-L)} = \sum_{\alpha}Y_{\Delta\left(B/3-L_{\alpha}\right)}$ and $c_{Bx}\equiv\sum_{y}J_{By}\left(J^{-1}\right)_{yx}$ with $J$ matrix defined in eq.~\eqref{eq:J matrix}. 
Assuming the EW symmetry is broken at $T_{\textrm{EW}}\sim160\,\textrm{GeV}$
above $T_{B-}$ \cite{DOnofrio:2014rug}, we have $c_{B(B-L)}=\frac{30}{97}$ and $c_{BY}=-\frac{7}{97}$ assuming top quarks are nonrelativistic (cf. ref. \cite{Harvey:1990qw}). Eq.~\eqref{eq:baryon_asymmetry} holds only down to temperature $T_{B-}$ below which the baryon number is frozen. If the Universe
is always hypercharge neutral $Y_{\Delta Y}=0$,
this implies that $U(1)_{B-L}$ has to be broken above $T_{B-}$ to generate a nonzero baryon asymmetry, as utilized in leptogenesis \cite{Fukugita:1986hr}, its variants \cite{AristizabalSierra:2007ur,AristizabalSierra:2009bh} and SO(10) baryogenesis \cite{Babu:2012iv}.~\footnote{Extending the SM by new fields which carry nonzero $B-L$ charges, baryogenesis can proceed with unbroken $U(1)_{B-L}$ \cite{Sierra:2013kba,Dick:1999je,Murayama:2002je,Gu:2012fg,Earl:2019wjw}. Some of this type of models can also accommodate the situation where compensating $B-L$ charge remains in the hidden sector and can serve as dark matter \cite{Gu:2012fg,Earl:2019wjw}.}
With $Y_{\Delta Y}=0$ and in the absence of new charges, if $(B-L)$-violating interactions remain in thermal equilibrium from $T_e$ down untill $T_{B-}$, baryogenesis will fail.~\footnote{See \cite{Nelson:1990ir,Deppisch:2013jxa,Deppisch:2015yqa,Deppisch:2017ecm} for studies to bound the scale of baryogenesis due to fast $(B-L)$-violating interactions.} 

It is usually required that $(B-L)$-violating interactions from new physics be out-of-equilibrium for a viable baryogenesis scenario (see for example refs. \cite{Fukugita:1990gb,Harvey:1990qw}). In this work, we will point out an orthogonal scenario. Rather, we argue that any \emph{in-equilibrium} $(B-L)$-violating interactions in fact facilitate baryogenesis and allow a new avenue of baryogenesis through out-of-equilibrium generation of asymmetry in the effective charges identified in eq. \eqref{eq:SM_global_symmetries}.

In general, fast $(B-L)$-violating interactions are more than welcome
since they will act as the source of nonzero $B-L$ charge in eq. (\ref{eq:baryon_asymmetry}) \cite{Fukugita:2002hu}.
The indication that $U(1)_{B-L}$ is broken from \emph{new physics} is ubiquitous.
If the SM is treated as an effective field theory at low energy, at
mass dimension-5, we have the Weinberg operator $\ell_{\alpha}H\ell_{\beta}H$
which breaks $B-L$ by two units and gives rise to Majorana neutrino
mass at low energy \cite{Weinberg:1979sa,Weinberg:1980bf}. It has
been verified that all dimension-6 \cite{Weinberg:1979sa,Weinberg:1980bf}
and dimension-8 operators \cite{Li:2020gnx} conserve $B-L$ while
for the 18 dimension-7 \cite{Lehman:2014jma,Liao:2016hru}
and 560 dimension-9 operators \cite{Li:2020xlh,Liao:2020jmn}, $B-L$ is violated by two units. If $\Delta B=-\Delta L=1$, they
lead to nucleon decay channels on top of those from the operators
that conserve $B-L$ \cite{Weinberg:1980bf}. If $\Delta L=2$, these operators contribute to Majorana neutrino mass and neutrinoless double beta decay processes \cite{Weinberg:1979sa,Weinberg:1980bf,Babu:2001ex,deGouvea:2007qla}
while if $\Delta B=2$, they can lead to neutron-antineutron oscillation
(see a review article \cite{Phillips:2014fgb}). Finally, a gauge $U(1)_{B-L}$
naturally arises from gauge symmetry $SO(10)$ in grand unified theory
and is broken spontaneously to the SM gauge group. 

If any of the $(B-L)$-violating processes discussed above are in thermal equilibrium in certain temperature regime, we can construct $B-L$ charge asymmetry as
\begin{eqnarray}
Y_{\Delta\left(B-L\right)} & = & \sum_{x}c_{\left(B-L\right)x}Y_{\Delta x},
\label{eq:B-L_asymmetry}
\end{eqnarray}
where $c_{(B-L)x} = \sum_y J_{(B-L)y} (J^{-1})_{yx}$ with $J$ matrix defined in eq.~\eqref{eq:J matrix}. 
At any range of temperature regime when $(B-L)$-violating interactions are in thermal equilibrium, a nonzero $Y_{\Delta(B-L)}$ is induced as long as any $x$ with nonzero coefficient $c_{(B-L)x}\neq 0$ has a nonvanishing charge $Y_{\Delta x} \neq 0$, which can be generated through out-of-equilibrium dynamics at the same range of temperature or much before (at higher temperature). As long as $(B-L)$-violating processes get out of equilibrium shortly after, the final baryon asymmetry will be given by (\ref{eq:baryon_asymmetry}). 

To recapitulate, this new class of baryogenesis can be realized by extending the SM with the following: 
\begin{itemize}
	\item some in-equilibrium $(B-L)$-violating interactions to enforce eq. \eqref{eq:B-L_asymmetry};
	\item some out-of-equilibrium processes that violate the effective symmetries in eq. \eqref{eq:SM_global_symmetries} and also provide sources of C and CP violation. In fact, all the global charges (besides $B-L$) are violated explicitly in the SM but new physics is required to have sufficient CP violation and out-of-equilibrium condition. 
\end{itemize}

As a concrete example, let us consider the temperature regime $10^{12}\,\textrm{GeV}\lesssim T\lesssim10^{14}\,\textrm{GeV}$
where the $(B-L)$-violating interactions mediated by the Weinberg operator $\frac{1}{\Lambda}\ell_{\alpha}H\ell_{\beta}H$ are in thermal equilibrium. 
The estimation goes as follows: with the neutrino mass $m_{\nu}=v^{2}/\Lambda$ where $v \equiv \left\langle H\right\rangle =174\,\textrm{GeV}$, this implies $\Lambda\sim3\times10^{14}\left(\frac{0.1\,\textrm{eV}}{m_{\nu}}\right)\,\textrm{GeV}$. Comparing the $(B-L)$-violating rate $\Gamma_{B-L}\sim m_{\nu}^{2}T^{3}/v^{4}$ for $T\lesssim\lambda^{2}\Lambda$ ($\lambda$ is some dimensionless coupling) to the Hubble rate $H=1.66\sqrt{g_{\star}}T^{2}/M_{\textrm{Pl}}$ ($g_{\star}=106.75$ for the SM and $M_{\textrm{Pl}} = 1.22 \times 10^{19}$ GeV), we obtain $T_{B-L}\sim10^{11}\left(\frac{0.1\,\textrm{ eV}}{m_{\nu}}\right)^{2}\,\textrm{GeV}$ above which the $(B-L)$-violating interactions are in equilibrium as long as $\lambda\gtrsim0.02\left(\frac{0.1\,\textrm{eV}}{m_{\nu}}\right)$.

Next, let us introduce some processes which violate the effective symmetries of the SM. In general, grand unified theories contain various such possibilities. For instance, the 126 Higgs in $SO(10)$ contains diquarks, dileptons and leptoquarks and their couplings to the SM fields violate several of the effective charges in eq. \eqref{eq:SM_global_symmetries}.
As an example, let us introduce a heavy $SU(2)_L$ singlet colored diquark $\psi$ with $B=2/3$, $Y=-2/3$ and the following decay channels $\psi \to dd,ss,bb$ which violate $U(1)_{u-d}$, $U(1)_{u-s}$ and $U(1)_{u-b}$
respectively by -2 units while respecting all other $U(1)_x$ symmetries (including $B$ and $L$).\footnote{For simplicity, we assume the decay channels are flavor diagonal.} (A nonzero charge will be induced in $U(1)_{\psi}$ during the genesis but this will go to zero once all $\psi$ particles have decayed away.) Assuming CPT invariance, the CP violation from the decays are related by
\begin{eqnarray}
\epsilon_{\psi \to dd}+\epsilon_{\psi \to ss}+\epsilon_{\psi \to bb} & = & 0,\label{eq:CPT_ex1}
\end{eqnarray}
where we have defined the CP parameter as $\epsilon_{\psi \to j}\equiv\frac{\Gamma\left(\psi\to j\right)-\Gamma\left(\overline{\psi}\to\overline{j}\right)}{\Gamma_\psi}$
with $\Gamma_\psi$ the total decay width of $\psi$ and $\Gamma\left(\psi \to j\right)$ and $\Gamma\left(\overline\psi \to \overline j\right)$ the partial decay widths.
The charge $Y_{\Delta x}$ generated from the decays of $\psi$ can be parametrized by
\begin{eqnarray}
Y_{\Delta x} & = & -2 \epsilon_{\psi(x)}\eta_{x}Y_{\psi}^{\textrm{eq}},
\end{eqnarray}
where $\psi(x)$ refers to the decay process which violates $x$ charge, $\eta_{x}\leq1$ is the efficiency for $x$ charge production through out-of-equilibrium dynamics and $Y_{\psi}^{\textrm{eq}}=n_{\psi}^{\textrm{eq}}/s$ with $n_{\psi}^{\textrm{eq}}$
the \emph{relativistic} equilibrium number density of $\psi$. 
If $\psi$ particles start from a thermal abundance and decay far from equilibrium when $T\ll m_{\psi}$, we have $\eta_{u-d}=\eta_{u-s}=\eta_{u-b}=1$. Making use of eqs.~\eqref{eq:B-L_asymmetry} and \eqref{eq:CPT_ex1} and with all conserved charges remain zero, after all $\psi$ particles have decayed above $T \sim 10^{12}\,\textrm{GeV}$, we end up with
\begin{eqnarray}
Y_{\Delta\left(B-L\right)} 
& = & \begin{cases}
\frac{1}{3}\epsilon_{\psi\to bb}Y_{\psi}^{\textrm{eq}} & T_{u}<T<T_{t}\\
\frac{9}{22}\epsilon_{\psi\to bb}Y_{\psi}^{\textrm{eq}} & T_{B}<T<T_{u}
\end{cases}.
\end{eqnarray}
Below $T_{B-L}$, $B-L$ is conserved and the final baryon asymmetry is given by eq. (\ref{eq:baryon_asymmetry}) with $Y_{\Delta Y} = 0$.
In order to obtain $Y_{\Delta B} \sim 10^{-10}$ in accordance with observation \cite{Zyla:2020zbs}, since $Y_{\psi}^{\textrm{eq}} \sim 10^{-3}$, one would need a reasonable CP violation of $\epsilon_{\psi \to bb} \sim 10^{-6}$.

\section{\label{sec:MSSM}The Minimal Supersymmetric SM}

In the MSSM, all the SM fermionic fields are promoted to superfields. For anomaly cancellation, two Higgs superfields $H_{u}$ and $H_{d}$ are introduced and we can choose the additional $U(1)_{PQ}$ conserved by all the superpotential terms~\footnote{Here, we denote  $\Psi=\left\{Q_{\alpha},U_{\alpha}^{c},D_{\alpha}^{c},\ell_{\alpha},E_{\alpha}^{c},H_u,H_d\right\} $ as the left-handed chiral superfields.} 
\begin{equation}
W = \mu H_u H_d + (y_u)_{\alpha\beta} Q_\alpha H_u U_\beta^c
+ (y_d)_{\alpha\beta} Q_\alpha H_d D_\beta^c
+ (y_e)_{\alpha\beta} \ell_\alpha H_u E_\beta^c,
\label{eq:superpotential}
\end{equation}
except $\mu H_{u} H_{d}$ with the following charge assignments
\begin{eqnarray}
q_{H_{d}}^{PQ} & = & q_{\ell_{\alpha}}^{PQ}=-\frac{q_{E_{\alpha}^{c}}^{PQ}}{2},q_{H_{u}}^{PQ}=-\frac{q_{E_{\alpha}^{c}}^{PQ}}{2}+3q_{D_{\alpha}^{c}}^{PQ},q_{Q_{\alpha}}^{PQ}=\frac{q_{E_{\alpha}^{c}}^{PQ}}{2}-q_{D_{\alpha}^{c}}^{PQ},q_{U_{\alpha}^{c}}^{PQ}=-2q_{D_{\alpha}^{c}}^{PQ}.
\end{eqnarray}
The mixed anomaly coefficients of $U(1)_{PQ}$ with $SU(3)_{c}$ and $SU(2)_L$ are respectively given by $A_{PQ33}=\frac{3}{2}\left(-3q_{D_{\alpha}^{c}}^{PQ}+q_{E_{\alpha}^{c}}^{PQ}\right)$ and $A_{PQ22}=-3q_{D_{\alpha}^{c}}^{PQ}+q_{E_{\alpha}^{c}}^{PQ}$.
The anomaly-free choice $-3q_{D_{\alpha}^{c}}^{PQ}+q_{E_{\alpha}^{c}}^{PQ}=0$
is the hypercharge and hence we will consider only the
solutions with $-3q_{D_{\alpha}^{c}}^{PQ}+q_{E_{\alpha}^{c}}^{PQ}\neq0$.
Since both $B$ and $L$ have the same anomaly coefficient $A_{B22}=A_{L22}=\frac{3}{2}$, an $SU(2)_{L}$ mixed anomaly-free charge can be formed \cite{Fong:2015vna}
\begin{eqnarray}
P & = & -\frac{3}{2}\frac{PQ}{-3q_{D^{c}}^{PQ}+q_{E^{c}}^{PQ}}+\frac{1}{c_{BL}}\left(c_{B}B+c_{L}L\right),
\end{eqnarray}
where $c_{BL}=c_{B}+c_{L}$ with $c_{B}$ and $c_{L}$ any numbers.
As for the $SU(3)_{c}$ mixed anomaly, we can cancel it with any of
the chiral symmetries of the quark fields $U(1)_{Q_{\alpha}}$, $U(1)_{U_{\alpha}^{c}}$
and $U(1)_{D_{\alpha}^{c}}$ with respective anomaly coefficients
$A_{Q_{\alpha}33}=1$ and $A_{U_{\alpha}^{c}33}=A_{D_{\alpha}^{c}33}=\frac{1}{2}$
(all the chiral charges are fixed to be 1). For instance, a completely
anomaly-free combination will be 
\begin{eqnarray}
\overline{P} & = & P+\frac{9}{2}u^{c}.\label{eq:Pbar}
\end{eqnarray}
Since $\overline{P}$ is violated explicitly only by
$\mu H_{u}H_{d}$ term, by comparing the interaction rate $\Gamma_{\overline P}\sim\mu^{2}/T$ to the Hubble rate $H=1.66\sqrt{g_{\star}}T^{2}/M_{\textrm{Pl}}$ ($g_{\star}=228.75$ for the MSSM), we obtain \cite{Ibanez:1992aj}
\begin{eqnarray}
T_{\overline{P}} & \sim & 2\times10^{7}\left(\frac{\mu}{100\,\textrm{GeV}}\right)^{2/3}.\label{eq:T_Pbar}
\end{eqnarray}
Above this temperature, $U(1)_{\overline{P}}$ is preserved by \emph{all} the MSSM interactions.

In a supersymmetric theory, gauginos can carry nonvanishing chemical
potentials and, scalar and fermionic components of a chiral
superfield do not necessarily carry the same chemical potentials.
This is captured by the $R$ symmetry in which the superspace
coordinate transforms as $\theta\to e^{i\phi}\theta$ where we fix
its $R$ charge to be 1. Requiring the superpotential in eq.~\eqref{eq:superpotential} to have an $R$ charge equals to 2, we have
\begin{eqnarray}
q_{H_{d}}^{R} & = & q_{\ell_{\alpha}}^{R}=2-\frac{3q_{D_{\alpha}^{c}}^{R}}{2},q_{H_{u}}^{R}=\frac{3q_{D_{\alpha}^{c}}^{R}}{2},q_{Q_{\alpha}}^{R}=\frac{q_{D_{\alpha}^{c}}^{R}}{2},q_{U_{\alpha}^{c}}^{R}=2-2q_{D_{\alpha}^{c}}^{R},q_{E_{\alpha}^{c}}^{R}=-2+3q_{D_{\alpha}^{c}}^{R}.
\end{eqnarray}
The $R$ symmetry only has $SU(2)_{L}$ mixed anomaly with anomaly
coefficient $A_{R22}=-1$ and an $SU(2)_{L}$ mixed anomaly-free
$R$ charge can be constructed \cite{Fong:2015vna}
\begin{eqnarray}
\overline{R} & = & R+\frac{2}{3c_{BL}}\left(c_{B}B+c_{L}L\right).\label{eq:Rbar}
\end{eqnarray}
Gaugino masses $m_{\widetilde{g}}$ break the $R$ symmetry explicitly and by comparing the associated interaction rate $\Gamma_{\overline R}\sim m_{\widetilde{g}}^{2}/T$
to the Hubble rate $H$, we obtain \cite{Ibanez:1992aj}
\begin{eqnarray}
T_{\overline{R}} & \sim & 8\times10^{7}\left(\frac{m_{\widetilde{g}}}{1\,\textrm{TeV}}\right)^{2/3}.\label{eq:T_Rbar}
\end{eqnarray}
Above this temperature, $U(1)_{\overline{R}}$ is preserved by \emph{all} the MSSM interactions. 

In any extension to the MSSM, one has the freedom choose $c_{B}$ and
$c_{L}$ such that $\overline{P}$ and/or $\overline{R}$ are conserved by the new interactions.~\footnote{R-parity-violating terms $U^c D^c D^c$, $Q D^c L$, $L L E^c$ conserve $P$ and $R$ while one can choose either $c_B=0$ or $c_L=0$ such that some of them conserve $\overline P$ and $\overline R$.} 
Let us consider a simple model of baryogenesis which breaks $\overline{P}$ and/or $\overline{R}$ without explicitly breaking $B$ and $L$.
We introduce a new chiral superfield $S$ uncharged under the
SM gauge symmetry with the following superpotential $\lambda SH_{u}H_{d}+\frac{1}{2}MSS$.
Taking $q_{S}^{\overline{P}}=0$ and $q_{S}^{\overline{R}}=1$, both $\overline{P}$
and $\overline{R}$ are broken by nonzero $\lambda$ respectively
by $\Delta\overline{P}=\frac{3}{2}$ and $\Delta\overline{R}=3-2=1$
while respecting all the $U(1)_x$ symmetries in eq. (\ref{eq:SM_global_symmetries}).
In this case, there is still an exactly conserved $R$ charge
\begin{eqnarray}
R_{c} & = & \overline{R}-\frac{2}{3}\overline{P}.
\end{eqnarray}
Nonzero charges can develop in both $\overline{R}$ and $\overline{P}$ (related by $R_c$) through CP-violating decays $S\to H_{u}H_{d}$.

Let us consider $B-L$ violation from the dimension-7 operator $D_{\alpha}^{c}U_{\beta}^{c}D_{\gamma}^{c}\ell_{\delta}H_{u}$
which can induce nucleon decay such as $n\to e^{-}\pi^{+}$. In order
to sufficiently suppress this process, the effective scale of the
operator should be of the order of $\gtrsim10^{11}\,\textrm{GeV}$ \cite{Weinberg:1980bf}.
With large couplings, the processes mediated by the operator can be
in equilibrium at temperature $10^{10}\,\textrm{GeV}\lesssim T\lesssim10^{11}$
GeV. We will take $c_{B}=-2c_{L}$ such that the operator
conserves $\overline{R}$.~\footnote{The Weinberg operator conserves $\overline{P}$ and $\overline{R}$ with the choice $c_{B}=-5c_{L}/3$.}
This operator violates $U(1)_{\overline{P}}$
and hence $R_{c}$ is no longer conserved. Assuming the operator involves
all generations of quarks, all effective symmetries related to quarks are violated.
With the remaining conserved charges ($e$, $\mu$, $Y$) being zero,
after all $S$ particles have decayed above $T \sim 10^{10}\,\textrm{GeV}$, we have
\begin{eqnarray}
Y_{\Delta\left(B-L\right)} & = & \frac{21}{187}Y_{\Delta\overline{R}}.
\end{eqnarray}
Below $T_{B-L}$, $B-L$ is conserved and the final baryon asymmetry will be given by eq. (\ref{eq:baryon_asymmetry}) with $Y_{\Delta Y} = 0$ and $c_{B(B-L)}=\frac{30}{97}$ assuming at $T_{B-}$, the thermal bath has the same relativistic degrees of freedom as in the SM.

\section{\label{sec:conclusions}Conclusions}

We have categorized all the effective $U(1)$ symmetries and their associate Noether charges in both the SM and the MSSM, 16 in the former and 18 in the latter. We have demonstrated that, together with fast $(B-L)$-violating interactions which are ubiquitous in the early Universe, and considering the effective symmetries in the SM or the MSSM above $T_e \sim 30$ TeV, asymmetries can be generated in any of the associated charges through processes that, can, but do not have to violate $B$ and/or $L$. 

Since this new avenue of baryogenesis calls for some in-equilbrium $(B-L)$-violating interactions in the early Universe, it generally implies an enhanced $\Delta(B-L)=2$ rate for experiments to observe: neutrinoless double beta decay ($\Delta L = 2$), nucleon decay ($\Delta B = -\Delta L = 1$) and neutron-antineutron oscillation ($\Delta B = 2$). A discovery of any of these phenomena will pinpoint the scale where $(B-L)$-violating interactions could be in thermal equilibrium and allow the identification of effective charges to realize baryogenesis above that scale. In the early Universe, $B-L$ violation could be the second crucial piece to the puzzle of cosmic baryon asymmetry after the SM baryon number violation.

Future direction in this exploration includes identifying new effective symmetries and explicit sources of $B-L$ violation that come out from more fundamental theories like grand unified theories. In studying these more specific models to realize the new baryogenesis proposed here, one will be able to have more definite predictions in the rate of $(B-L)$-violating processes.

\vspace{1cm}

\noindent \textbf{Note added}: While this work is being written up, a similar idea appears on arXiv \cite{Domcke:2020quw}, which considers the dynamics of type-I seesaw as the source of $B-L$ violation. Our results, where there are overlaps, are consistent with each other.

\section{Acknowledgments}
C.S.F. acknowledges the support by FAPESP grant 2019/11197-6 and CNPq grant 301271/2019-4. He is grateful to the support of family members and colleagues, near and far, during this Covid-19 isolation. He would like to thank Shaikh Saad for the comments on the manuscript.

\appendix

\section{\label{app:symmetries_T}$U(1)$ breaking in the SM and MSSM}

Firstly, we would like to estimate the temperatures in which baryon flavor
numbers $U(1)_{B_{\alpha}}$ and quark flavors $U(1)_{q}$ with $q=\left\{ u,d,c,s,t,b\right\} $
are violated. All chiral symmetries $U(1)_{q}$ are broken by the
$SU(3)_{c}$ instanton-induced effective operator ${\cal O}_{SU(3)}\sim\prod_{\alpha}\left(QQU^{c}D^{c}\right)_{\alpha}$. The interactions induced by this operator get into thermal equilibrium at $T_{u}\sim2\times10^{13}\,\textrm{GeV}$ \cite{Garbrecht:2014kda}. It is then convenient to consider the following combinations of $U(1)_{x}$ with
\begin{eqnarray}
x & = & \left\{ B,B_{3}-B_{2},B_{3}+B_{2}-2B_{1},t,u,u-b,\text{\ensuremath{\tau}},u-c,\mu,u-s,u-d\right\} .
\end{eqnarray}
All the charges besides $t$, $u$ and $B$ are free from $SU(2)_L$ and $SU(3)_c$ mixed anomalies. $t$ is the first to be broken at $T_t \sim 10^{15}$ GeV by top Yukawa mediated interactions, $u$ is broken at $T_u$ by ${\cal O}_{SU(3)}$ while $B$ is broken at $T_{B}\sim2\times10^{12}\,\textrm{GeV}$ by ${\cal O}_{SU(2)}\sim\prod_{\alpha}\left(QQQ\ell\right)_{\alpha}$ \cite{Garbrecht:2014kda}.

Next we will estimate when the rest of the symmetries are broken as cosmic temperature decreases. In order to do so, we first have to determine
the appropriate basis of quarks in the thermal bath. The general quark Yukawa interactions are given by
\begin{eqnarray}
-{\cal L}_{Y} & = & \overline{Q}Y_{u}U\epsilon H^{*}+\overline{Q}Y_{d}DH.
\end{eqnarray}
In the thermal bath, all the quarks would acquire chirality-conserving
thermal mass due to interactions and one should consider the thermal mass basis. The contributions from gauge interactions are flavor blind while those from the Yukawa interactions are $m_{U}^{2}/T^2 \propto\left(Y_{u}^{\dagger}Y_{u}\right)$,
$m_{D}^{2}/T^2 \propto\left(Y_{d}^{\dagger}Y_{d}\right)$ and $m_{Q}^{2}/T^2 \propto\left(Y_{u}Y_{u}^{\dagger}\right)+\left(Y_{d}Y_{d}^{\dagger}\right)$.
In general, the Yukawa couplings can be diagonalized as follows $\hat{Y}_{u}=U_{u}Y_{u}V_{u}^{\dagger}\equiv\textrm{ diag}\left(y_{u},y_{c},y_{t}\right)$
and $\hat{Y}_{d}=U_{d}Y_{d}V_{d}^{\dagger}\equiv\textrm{ diag}\left(y_{d},y_{s},y_{b}\right)$
where $U_{u,d}$ and $V_{u,d}$ are unitary matrices and $(y_u,y_c,y_t,y_d,y_s,y_b)(174\,\textrm{GeV})$ can be identified with the physical masses of the quarks. In the basis $U'=V_{u}U$ and $D'=V_{d}D$, both $m_{U}^{2}$ and $m_{D}^{2}$ are diagonalized while 
\begin{eqnarray}
m_{Q}^{2}/T^2 & \propto & U_{u}^{\dagger}\left(\hat{Y}_{u}^{2}+V_{\textrm{CKM}}\hat{Y}_{d}^{2}V_{\textrm{CKM}}^{\dagger}\right)U_{u},
\end{eqnarray}
where $V_{\textrm{CKM}}=U_{u}U_{d}^{\dagger}$ is identified with the Cabibbo-Kobayashi-Maskawa mixing matrix. The fact that $V_{\textrm{CKM}}\neq I_{3\times3}$ results in the breaking
of $U(1)_{B_\alpha}$. By going to a basis $Q'=U_{Q}Q$, $m_{Q}^{2}$
can be diagonalized as well. We can split $V_{\textrm{CKM}}=I_{3\times 3}+\delta V$ and $U_{Q}=U_{u}+\delta U$. Since the elements of $\delta V$ are in general much smaller than unity (the largest elements being $\delta V_{12}\sim\delta V_{21}\sim0.23$),
we can solve for $\delta U$ perturbatively. At the leading order, we
obtain
\begin{eqnarray}
\left[\delta UU_{u}^{\dagger}\right]_{mn} & = & -\delta V_{mn}\frac{\left(\hat{Y}_{d}\right)_{m}^{2}-\left(\hat{Y}_{d}\right)_{n}^{2}}{\left(\hat{Y}_{u}+\hat{Y}_{d}\right)_{m}-\left(\hat{Y}_{u}+\hat{Y}_{d}\right)_{n}},\;\;\;\;\;\left(m\neq n\right).
\end{eqnarray}

In the thermal mass basis where $m_U^2$, $m_D^2$ and $m_Q^2$ are all diagonal, we have
\begin{eqnarray}
{\cal L}_{Y} & = & \overline{Q'}\hat{Y}_{u}U'\epsilon H^{*}+\overline{Q'}\hat{Y}_{d}D'H
+\overline{Q'}y_{Qu}U'\epsilon H^{*}+\overline{Q'}y_{Qd}\hat{Y}_{d}D'H,
\end{eqnarray}
where for the last two terms which violate $U(1)_{B_\alpha}$, we
have denoted $y_{Qu}\equiv\delta UU_{u}^{\dagger}\hat{Y}_{u}$ and
$y_{Qd}\equiv\left(\delta V+\delta UU_{u}^{\dagger}\right)\hat{Y}_{d}$.
Keeping only the leading terms considering $y_{u}<y_{d}<y_{s}<y_{c}<y_{b}<y_{t}$,
we have
\begin{eqnarray}
y_{Qu} & = & \left(\begin{array}{ccc}
0 & -\delta V_{12}y_{s}^{2} & -\delta V_{13}y_{b}^{2}\\
-\delta V_{21}\frac{y_{s}^{2}}{y_{c}}y_{u} & 0 & -\delta V_{23}y_{b}^{2}\\
-\delta V_{31}\frac{y_{b}^{2}}{y_{t}}y_{u} & -\delta V_{32}\frac{y_{b}^{2}}{y_{t}}y_{c} & 0
\end{array}\right),\\
y_{Qd} & = & \left(\begin{array}{ccc}
0 & \left(1-\frac{y_{s}^{2}}{y_{c}}\right)\delta V_{12}y_{s} & \left(1-\frac{y_{b}^{2}}{y_{t}}\right)\delta V_{13}y_{b}\\
\left(1-\frac{y_{s}^{2}}{y_{c}}\right)\delta V_{21}y_{d} & 0 & \left(1-\frac{y_{b}^{2}}{y_{t}}\right)\delta V_{23}y_{b}\\
\left(1-\frac{y_{b}^{2}}{y_{t}}\right)\delta V_{31}y_{d} & \left(1-\frac{y_{b}^{2}}{y_{t}}\right)\delta V_{32}y_{s} & 0
\end{array}\right).
\end{eqnarray}

For the rate of quark-Yukawa-coupling-mediated interactions, we will use the result
of \cite{Garbrecht:2014kda} $\Gamma_y \approx10^{-2}c\left(T\right)y^{2}T$
where $y$ refers to elements of $\hat{Y}_{u}$, $\hat{Y}_{d}$, $y_{Qu}$ and $y_{Qd}$ and we will make an extrapolation in $c(T)$ to take into account the running of strong coupling. We will also consider the running of quark Yukawa couplings \cite{Xing:2007fb} but ignore the running of mixing angles. 

Next, we define the transition temperature $T_{x}$ as when $U(1)_x$ is first broken at $\Gamma_{x}\left(T_{x}\right)=H(T_{x})$ where the Hubble rate is $H(T)=1.66\sqrt{g_{\star}}T^{2}/M_{\textrm{Pl}}$ ($g_\star = 106.75$ for the SM).
We estimate that $U(1)_{B_{3}}$ and $U(1)_{B_{2}}$ are both broken
first by $\left(y_{Qd}\right)_{23}$ at $T_{B_{3}-B_{2}}\sim 9 \times 10^{8}\,\textrm{GeV}$
while $U(1)_{B_{1}}$ is broken first by $\left(y_{Qd}\right)_{12}$
at $T_{B_{3}+B_{2}-2B_{1}}\sim10^{7}\,\textrm{GeV}$. Similarly, from
$\hat{Y}_{u}$ and $\hat{Y}_{d}$, we estimate $T_{t} \sim 10^{15}\,\textrm{GeV}$, $T_{u-b} \sim 3\times10^{11}\,\textrm{GeV}$, 
$T_{u-c} \sim 2\times10^{10}\,\textrm{GeV}$, $T_{u-s} \sim 3\times10^{8}\,\textrm{GeV}$ and $T_{u-d} \sim 2\times10^{6}\,\textrm{GeV}$. 

For completeness, we will also estimate the rate of charged-lepton-Yukawa-coupling-mediated interactions with $\Gamma_y \approx 5\times 10^{-3} y^{2}T$ from ref. \cite{Garbrecht:2014kda}. For convenience of the readers, we collect here the $T_x$ in the order when $U(1)_x$ is broken as we go down in cosmic temperature: 
\begin{eqnarray}
T_{t} & \sim & 10^{15}\,\textrm{GeV},\nonumber\\
T_{u} & \sim & 2\times10^{13}\,\textrm{GeV},\nonumber\\
T_{B} & \sim & 2\times10^{12}\,\textrm{GeV},\nonumber\\
T_{\tau} & \sim & 4\times10^{11}\,\textrm{GeV},\nonumber\\
T_{u-b} & \sim & 3\times10^{11}\,\textrm{GeV},\nonumber\\
T_{u-c} & \sim & 2\times10^{10}\,\textrm{GeV},\\
T_{\mu} & \sim & 10^{9}\,\textrm{GeV}\nonumber\\
T_{B_{3}-B_{2}} & \sim & 9 \times 10^{8}\,\textrm{GeV},\nonumber\\
T_{u-s} & \sim & 3\times10^{8}\,\textrm{GeV},\nonumber\\
T_{B_{3}+B_{2}-2B_{1}} & \sim & 10^{7}\,\textrm{GeV},\nonumber\\
T_{u-d} & \sim & 2\times10^{6}\,\textrm{GeV},\nonumber\\
T_{e} & \sim & 3\times10^{4}\,\textrm{GeV}.\nonumber
\end{eqnarray}
All the symmetries above are broken at $T_x$ and not restored again with the exception of $U(1)_B$ which is restored as a good symmetry below $T_{B-} \sim 130\,\textrm{GeV}$.
The symmetries not in the list: $U(1)_{B/3-L_\alpha}$ are never broken in the SM nor the MSSM while the hypercharge $U(1)_Y$ is spontaneously broken at $T_{\textrm{EW}} \sim 160\,\textrm{GeV}$. The smooth transition between the regime can be described in a unified manner in the density matrix formalism \cite{Raffelt:1992uj,Sigl:1992fn,Blanchet:2008hg,Blanchet:2011xq} and will be explored in an upcoming publication. In the MSSM, besides modification to the running couplings, and the change of relativistic degrees of freedom, the transition temperatures for down type quark and charged lepton will be modified by an overall factor of $1+\tan^{2}\beta$ where $\tan\beta \equiv \left<H_u\right>/\left<H_d\right>$.

If $(B-L)$-violating processes are in thermal equilibrium in certain temperature regime, we can construct $B-L$ charge as~\footnote{The three charges $B/3-L_\alpha$ can be rewritten in another basis $B-L$, $L_1-L_2$ and $L_1-L_3$. Only $B-L$ charge contributes to the final baryon asymmetry.}
\begin{eqnarray}
Y_{\Delta\left(B-L\right)} & = & \sum_{x}c_{\left(B-L\right)x}Y_{\Delta x},
\label{eq:B-L_asymmetry}
\end{eqnarray}
where the explicit coefficients are collected in Table \ref{tab:SM_B-L_coeff}.

\begin{table}
	\begin{tabular}{|c|c|c|c|c|c|c|c|c|c|c|c|c|c|}
		\hline 
		$T_x/c_{(B-L)x}$ & $t$ & $u$ & $B$ & $\tau$ & $u-b$ & $u-c$ & $\mu$ & $B_{3}-B_{2}$ & $u-s$ & $B_{3}+B_{2}-2B_{1}$ & $u-d$ & $e$ & $Y$\tabularnewline
		\hline 
		\hline 
		& $-\frac{3}{5}$ & $\frac{3}{5}$ & $\frac{2}{5}$ & $\frac{1}{5}$ & $-\frac{3}{5}$ & $\frac{3}{5}$ & $\frac{1}{5}$ & $0$ & $\frac{3}{5}$ & $0$ & $\frac{3}{5}$ & $\frac{1}{5}$ & $\frac{6}{5}$\tabularnewline
		\hline 
		$T_{t}$ & $0$ & $\frac{2}{3}$ & $\frac{1}{3}$ & $0$ & $-\frac{2}{3}$ & $\frac{1}{2}$ & $0$ & $-\frac{1}{4}$ & $-\frac{1}{2}$ & $-\frac{1}{12}$ & $-\frac{1}{2}$ & $0$ & $1$\tabularnewline
		\hline 
		$T_{u}$ & $0$ & $0$ & $\frac{23}{44}$ & $-\frac{1}{22}$ & $-\frac{6}{11}$ & $\frac{27}{44}$ & $-\frac{1}{22}$ & $-\frac{27}{88}$ &  $-\frac{15}{44}$ & $-\frac{9}{88}$ & $-\frac{15}{44}$ & $-\frac{1}{22}$ & $\frac{21}{22}$\tabularnewline
		\hline 
		$T_{B}$ & $0$ & $0$ & $0$ & $\frac{6}{29}$ & $-\frac{20}{29}$ & $\frac{45}{58}$ & $\frac{6}{29}$ & $-\frac{45}{116}$ & $-\frac{25}{58}$ & $-\frac{15}{116}$ & $-\frac{25}{58}$ & $\frac{6}{29}$ & $\frac{35}{29}$\tabularnewline
		\hline 
		$T_{\tau}$ & $0$ & $0$ & $0$ & $0$ & $-\frac{2}{3}$ & $\frac{3}{4}$ & $\frac{1}{6}$ & $-\frac{3}{8}$ & $-\frac{5}{12}$ & $-\frac{1}{8}$ & $-\frac{5}{12}$ & $\frac{1}{6}$ & $\frac{7}{6}$\tabularnewline
		\hline 
		$T_{u-b}$ & $0$ & $0$ & $0$ & $0$ & $0$ & $\frac{1}{2}$ & $0$ & $0$ & $-\frac{1}{2}$ & $0$ & $-\frac{1}{2}$ & $0$ & $1$\tabularnewline
		\hline 
		$T_{u-c}$ & $0$ & $0$ & $0$ & $0$ & $0$ & $0$ & $-\frac{1}{9}$ & $\frac{1}{6}$ & $-\frac{4}{9}$ & $-\frac{1}{6}$ & $-\frac{2}{9}$ & $-\frac{1}{9}$ & $\frac{8}{9}$\tabularnewline
		\hline 
		$T_{\mu}$ & $0$ & $0$ & $0$ & $0$ & $0$ & $0$ & $0$ & $\frac{21}{124}$ & $-\frac{14}{31}$ & $-\frac{21}{124}$ & $-\frac{7}{31}$ & $-\frac{3}{31}$ & $\frac{28}{31}$\tabularnewline
		\hline 
		$T_{B_{3}-B_{2}}$ & $0$ & $0$ & $0$ & $0$ & $0$ & $0$ & $0$ & $0$ & $-\frac{56}{141}$ & $-\frac{7}{47}$ & $-\frac{35}{141}$ & $-\frac{5}{47}$ & $\frac{42}{47}$\tabularnewline
		\hline 
		$T_{u-s}$ & $0$ & $0$ & $0$ & $0$ & $0$ & $0$ & $0$ & $0$ & $0$ & $0$ & $-\frac{7}{17}$ & $-\frac{3}{17}$ & $\frac{14}{17}$\tabularnewline
		\hline 
		$T_{B_{3}+B_{2}-2B_{1}}$ & $0$ & $0$ & $0$ & $0$ & $0$ & $0$ & $0$ & $0$ & $0$ & $0$ & $-\frac{7}{17}$ & $-\frac{3}{17}$ & $\frac{14}{17}$\tabularnewline
		\hline 
		$T_{u-d}$ & $0$ & $0$ & $0$ & $0$ & $0$ & $0$ & $0$ & $0$ & $0$ & $0$ & $0$ & $-\frac{3}{10}$ & $\frac{7}{10}$\tabularnewline
		\hline 
		$T_{e}$ & $0$ & $0$ & $0$ & $0$ & $0$ & $0$ & $0$ & $0$ & $0$ & $0$ & $0$ & $0$ & $\frac{8}{11}$\tabularnewline
		\hline 
	\end{tabular}
	
	\caption{The coefficients $c_{\left(B-L\right)x}$ in eq. (\ref{eq:B-L_asymmetry}) for the SM assuming $(B-L)$-violating interactions are in thermal equilibrium in various temperature regime. The first column indicates the temperature $T_x$ where $U(1)_x$ is first broken while the first row indicates the label $x$ of the coefficients $c_{(B-L)x}$. \label{tab:SM_B-L_coeff}}
\end{table}

\bibliography{main}

\end{document}